\documentclass[showpacs,showkeys,pra,twocolumn]{revtex4-2}
\usepackage{amsmath}
\usepackage{graphicx}
\usepackage{color}
\usepackage{amssymb}
\usepackage{float}
\begin{document}
\title{Dynamics of bright solitons in spin-orbit coupled Bose-Einstein condensates under the influence of optical and Rabi-coupling lattices}
\author{Sumaita Sultana}
%\email{reach2sumaita@gmail.com}
\affiliation{Department of Physics, Kazi Nazrul University, Asansol-713340, W.B., India}
\author{Sk Siddik}
%\email{amanandsafin@gmail.com}
\affiliation{Department of Physics, Kazi Nazrul University, Asansol-713340, W.B., India}
\author{Golam Ali Sekh}
\email{skgolamali@gmail.com}
\affiliation{Department of Physics, Kazi Nazrul University, Asansol-713340, W.B., India}
\begin{abstract}
We consider coupled matter-wave bright solitons in spatial coherence spin-orbit coupled Bose-Einstein condensates in optical and Rabi-coupling lattice potentials and find an effective potential for separation of coupled matter-wave solitons. We show that the Rabi lattice significantly affects the potential near the region of overlap between the solitons. We study the effects of Rabi lattice on the dynamics of solitons for different initial overlap and that the dynamics is controlled by the interplay between the Rabi lattice and inter-component interaction. We investigate that, by tuning the Rabi lattice strength, one can realize localize, periodic and splitting dynamics of the total density profile of the coupled solitons. 
\end{abstract}
\maketitle
\section{Introduction}
Realization of spin-orbit coupled Bose-Einstein condensates through the generation artificial gauge potential receives interest  in atomic physics for its fundamental importance and in condensed matter physics for the possibility to simulate different phenomena in a tunable way \cite{r1,r2,r3}. Experimental flexibility to tune different parameters  makes it possible to  observe and propagate matter-wave bright solitons in spin-orbit coupled Bose-Einstein condensates (SOC-BECs). Certainly, it  reveals the nonlinear aspect of the system \cite{r4,zhu1}. In SOC-BECs,  the Galilean invariance is broken and the propagation of soliton is disturbed. This results changes in the shape of matter-wave solitons  during time evolution. Particularly, the soliton profile gets several peaks during its propagation and the number of peaks changes with a slight change of its velocity \cite{xu,gar1,gar0}. {The SOC-BEC with weak Raman coupling in the spatiotemporal  reversal parity-time  symmetric potentials holds oscillatory non-degenerate solitons and the collision between two nondegenerate solitons is  elastic \cite{gar1a}. In addition, the system support high amplitude peregrine solitons \cite{gar1aa}, reduces the life-time of a ring dark soliton \cite{gar1bb} and generates abundant vortex \cite{gar1cc}. }

In a spin-orbit coupled BEC, a proper combination of radio-frequency magnetic field and optical-Raman coupling field can induce a spin-dependent lattice, called Zeeman lattice (ZL)\cite{gar2}. 
In bosonic system, the ZL were first introduced by Jimenez-Garcia et al.\cite{gar}. The generation of ZL directly contrasts the optical lattice (OL) created with optical standing waves\cite{malo}. Particularly, OL lattices are same for both the components while  the ZL have opposite sign in two components. In the OL,  atoms acquire a quantum mechanical phase as it hops from one site to other.   In presence of the ZL, the Galilean invariance of the system is recovered. This virtue of the ZL is exploited in the context of Bloch oscillation in \cite{ome} and demonstrated that ZL  can be used to accelerate the atom. This lattice also results  periodic doubling of Bloch oscillations due to band crossing at the edge of Brillouin zone \cite{karta}  and provides  a platform for the prediction of novel spin states\cite{26a}. Properties of fundamental gap soliton in ZL with defect and multipole gap solitons in parity-time-symmetric Zeeman lattices have be studied in \cite{konotop,zhu} and \cite{xing} respectively. 

Our objective in this work is to  study dynamics of  matter-wave bright solitons in  spin-orbit coupled Bose-Einstein condensates where long range coherent exists between different spin components \cite{putra}. { More specifically, we examine the effects of potential induced by periodically modulating Rabi coupling term, called Rabi-coupling lattice (RL), on the coupled solitons embedded in the OL. Here  the RL have same sign in both the components but its effects comes through  linear coupling.}  The RL affects the effective potential of the coupled solitons embedded in OL and thus the dynamics of matter-wave bright solitons. In the region of stronger overlap between the component solitons, the effective potential is significantly modified if RL strength is relative large. Here the inter-component interaction inter-plays with the RL for creating local minima which can support stable coupled solitons. In particular, we see that, for appropriately chosen initial environment in the effective potential, the total density profile of the coupled solitons can be localized in the principle minimum of the effective potential or it can be splited into two parts and move to the nearest local  minima of the effective potential. Depending on the  strength of RL, the coupled solitons   get shifted to different spatial location of the system even without splitting due to the change of local energy minimum arising from inter-component and RL potentials. We see that, this is quite similar to the dynamical self-trapping where the dynamical localization is described as a result of change of nonlinear lattice minima with time \cite{go3}.

In section II, we introduce a theoretical model based on variational approach for the bright solitons in spatially coherent SOC-BECs and find an effective potential for the separation of the soliton pair. In a binary BEC without spin-orbit coupling, the effective potential for the separation are found very effective to describe the dynamics of coupled solitons in presence of spin-independent optical lattices \cite{go1,go2}. Here we extend the model to deal with Rabi-coupled lattice. We study the dynamics of the coupled soliton  numerically using the methods of lines (MOL) for different initial environments in the effective lattice potential in section III. Finally, in section IV, we make concluding remarks.

\section{Theoretical formulation}
{We consider a  spin-orbit coupled (SOC) Bose-Einstein condensate which is equivalent to a SOC electronic system with equal Rasba and Dresselhaus couplings and an uniform magnetic field acts in the $x-z$ plane \cite{gar1,zhu1}. In this system, we introduce a period trapping potential for each component, known as optical lattice,  and a periodic potential by spatially modulating Rabi coupling, termed as Rabi-coupling lattice (RL).}
The spin-orbit coupled Bose-Einstein condensate with optical and Rabi-coupled lattice potentials is modeled by the following Gross-Pitaevskii equations \cite{konotop,zhu,zhu1}.
\begin{eqnarray}
\!\!\!\!\!\! i\frac{\partial \psi_1}{\partial t}\!=\!-\frac{1}{2}\frac{\partial^2\psi_1}{\partial x^2}\!+\!i k_s \frac{\partial \psi_1}{\partial x}\!+\!V (x)\psi_1\!+\!\Omega(x)\psi_2\!+\!\Gamma_{12}\psi_1,
\label{eq1}\\
%%%%%%%%%%%%%%%%%%%%%
\!\!\!\!\!\!i\frac{\partial \psi_2}{\partial t}\!=\!-\frac{1}{2}\frac{\partial^2\psi_2}{\partial x^2}\!-\!i k_s \frac{\partial \psi_2}{\partial x}\!+\!V(x)\psi_2\!+\!\Omega(x)\psi_1\!+\!\Gamma_{21}\psi_2.
\label{eq2}
\end{eqnarray}
with $\Gamma_{ij}=(g |\psi_i|^2+g_{12}|\psi_j|^2)$.
Here $\psi_1$ and $\psi_2$ are the order parameters for the first and second components. The parameters $k_s$, $g_j$ and $g_{12}$ represent respectively strengths of spin-orbit coupling, inter-atomic and intra-atomic interactions.  The periodic potentials $V_j=V_0\cos(2 k x)$ and $\Omega(x)=\Omega_0\cos(k_m x)$ represent respectively optical lattice and  Rabi-coupling lattice. In Eqs. (\ref{eq1})  and (\ref{eq2})   the units of length, time and energy are $1/k_{R}$, $m/(\hbar k^2_{R})$ and $\hbar^2 k_R^2/m$ with $k_R$, wavenumber of the Raman laser beams. 

Lagrangian density for  Eqs. (\ref{eq1})  and (\ref{eq2}) is given by
\begin{eqnarray}
{\cal L} &=&\sum_{j=1,2}\frac{1}{2}\left|\frac{\partial \psi_j}{\partial x}\right|^2+i(-)^j k_s\psi_j^*\frac{\partial \psi_j}{\partial x}+V_j(x)|\psi_j|^2\nonumber\\&+&
\left(g|\psi_{j}|^2+g_{12}|\psi_{|j-3|}|^2\right)|\psi_{j}|^2+\frac{i}{2}\left(\psi_j\frac{\partial \psi_j^*}{\partial t}-\psi_j^*\frac{\partial \psi_j}{\partial t}\right)\nonumber\\&+&
\Omega(x) \psi_j^*\psi_{|j-3|}.
\label{eq3}
\end{eqnarray}
{Spin-orbit coupled BEC is a pseudo two-component system differing in spins. These two spin components are interacting and they are spatially coherent and this coherence is maintained even the spatial overlap is small implying the existence long-range coherence \cite{putra}. Therefore, in the weak coupling limit and, in absence of optical lattices, we approximate Gaussian shaped trial solution for Eqs. (\ref{eq1}) and (\ref{eq2}) as follows.} 
\begin{eqnarray}
\psi_j(x,t)&=&\sqrt{\frac{N_j}{\sqrt{\pi} a_j(t)}} \exp[-(x+(-1)^j x_j(t))^2/(2 a_j(t))^2 \nonumber\\&+& i \dot{x}_j(t)(x+(-1)^j x_j(t))+i \phi_j(t)].
\label{eq4}
\end{eqnarray}
Here $j=1,2$ and $x_j(t)=x_0(t)/2$. In Eq. (\ref{eq4}), $N_j$, $a_j$, $\phi_j$ and $x_j$ are the variational parameters and represent respectively, norm, width, phase and center of mass of the soliton respectively. Note that Eqs.(\ref{eq1}) and (\ref{eq2}) give bright solution \cite{zhu1} and the Gaussian-shaped trial solution is found suitable for bright soliton \cite{go4}. {Understandably, it is a chirp free trial solution and can accurately describe the dynamics of center-of-mass of coupled solitons in two-component Bose-Einstein condensates \cite{go1,go2}.} 

To obtain equations for the parameters, we write averaged Lagrangian
\begin{eqnarray}
\langle {\cal L}\rangle &=&\sum_j(\frac{N_j}{4 a_j^2}+\frac{g N_j^2}{2\sqrt{2 \pi}a_j}+ \frac{V_0 a_j N_j}{\exp(k^2 a_j^2)}\cos(k x_0)+N_j\dot{\phi}_j\nonumber\\&-&\frac{1}{8}N_j \dot{x}_0^2-\frac{k_s}{2} N_j\dot{x}_0)\!+\!{\frac{g_{12} N_{1} N_{2} e^{-\frac{x_0^2}{a_{1}^2+a_{2}^2}}}{\sqrt{\pi } \sqrt{a_{1}^2+a_{2}^2}}}\!+\!\left<{\cal L}_z\right>.
\label{eq5}
\end{eqnarray}
where 
\begin{eqnarray}
&&\left<{\cal L}_z\right>=\,\,\frac{2 \Omega_0}{w}\sqrt{2 N_1 N_2 a_1 a_2}\,\,  e^{-2(k_m^2+x_0^2)-a_1^2 a_2^2 \dot{x}_0^2}\!\times \nonumber\\&& \!\!\!\!e^{-(k_m-(-1)^jx_0)^2/(2 w^2)}\cos\left(\!\!\phi+\frac{s^2((-1)^jk_m+x_0)\dot{x_0}}{2w^2}\right).\,\,\,\,\,\,\,\,\,
\label{eq6}
\end{eqnarray}
Here $\phi=\phi_1-\phi_2$, $w=\sqrt{a_1^2+a_2^2}$ and $s=\sqrt{a_1^2-a_2^2}$.

{Equations for the variational parameters are obtained from the vanishing condition of the variational derivative $\frac{\delta \langle {\cal L}\rangle}{\delta y_j}=\frac{d}{dt}\left( \frac{\partial \langle {\cal L}\rangle}{\partial\dot{y}_j}\right)-\frac{\partial \langle {\cal L}\rangle}{\partial{y}_j}$, where $y_j$ stands for variational parameter and $\dot{y}_j=\frac{d y_j}{dt}$.} For $y_j=\phi_j$ we can write
\begin{eqnarray}
\frac{d}{dt}\left( N_1+N_2\right)=0.
\end{eqnarray}
% From $\delta \langle {\cal L}\rangle/\delta A_j=0 $ we get
% 
%\begin{eqnarray}
%\frac{N_j}{2 a_j^2}&+&g_{11}\frac{\sqrt{2}N_j^2}{a_j}+\frac{2g_{12}}{\sqrt{\pi} w(t)} N_1 N_2 e^{-x_0^2/w^2}+2V_0N_j e^{-k^2 a_j^2/4}\cos(k x_0/2)-(k_s {\dot{x}}_0+\frac{{\dot{x}_0}^2}{4})N_j+2 N_j{\dot{\Phi}}_j\nonumber\\&+&
%A_j\frac{\partial\langle {\cal L}_z\rangle}{\partial A_j}=0
%\end{eqnarray}
{This implies that the total norm $N=N_1+N_2$ of the binary system is conserved.} For $\delta \langle {\cal L}\rangle/\delta a_j=0 $, we get
\begin{eqnarray}
&-&{\frac{N_j}{2 a_j^3}}+\frac{g_{j} N_j^2}{2 \sqrt{2\pi} a_j^2}-{2V_0 N_1}k^2\,e^{-k^2 a_1^2}\cos(k x_0)\nonumber\\&+&\frac{g_{12} a_j N_1 N_2}{\sqrt{\pi}w^5}(w^2 -2 x_0^2)\,e^{-x_0^2/w^2}+\frac{\partial\langle {\cal L}_z\rangle}{\partial a_j}=0.
\end{eqnarray}

From $\delta \langle {\cal L}\rangle/\delta x_0=0 $ we get
\begin{eqnarray}
&&\frac{N}{4}\frac{d^2 x_0}{dt^2}-{V_0 k}\left(N_1 e^{-k^2 a_1^2}+N_2 e^{-k^2 a_2^2}\right)\sin(k x_0)\nonumber\\&-&\frac{2 g_{12} N_1 N_2 x_0}{\sqrt{\pi}w^3} e^{-x_0^2/w^2}-\frac{\partial\langle {\cal L}_z\rangle}{ \partial x_0}=0.
\label{eq9}
\end{eqnarray}
Note that, in writing Eq.(\ref{eq9}), we neglect all the terms containing second order-derives of $\langle {\cal L}_z\rangle$ with respect to different parameters. Therefore, Eq.(\ref{eq9}) can be used to get qualitative idea on the  dynamics of soliton pair in presence of Rabi-coupling lattice.
% and this approximation leads 
%\ddot{x}_0\frac{\partial^2 \langle {\cal L}_z\rangle}{\partial \dot{x}_0^2}+\dot{x}_0\frac{\partial^2 \langle {\cal L}_z\rangle}{\partial \dot{x}_0 \partial x_0}+\dot{\phi}_1\frac{\partial^2 \langle {\cal L}_z\rangle}{\partial \dot{x}_0 \partial \phi_1}+\dot{\phi}_2\frac{\partial^2 \langle {\cal L}_z\rangle}{\partial \dot{x}_0 \partial \phi_2}+\dot{a}_1\frac{\partial^2 \langle {\cal L}_z\rangle}{\partial \dot{x}_0 \partial a_1}+\dot{a}_2\frac{\partial^2 \langle {\cal L}_z\rangle}{\partial \dot{x}_0 \partial a_2}+\frac{\partial\langle {\cal L}_z\rangle}{ \partial x_0}\approx \frac{\partial\langle {\cal L}_z\rangle}{ \partial x_0}$.

\subsection{Effective potential for the separation of soliton pair}
We have noted that the coherence between the components of the spin-orbit coupled Bose-Einstein condensates exists even they spatially separated.  Considering the fact that the spin flipping between components is negligible  and thus population imbalance is either zero or remains constant with time. From Eq. (\ref{eq9}), we write the following effective potential for the separation.
{\begin{eqnarray}
V_{\rm eff}&=&k^2\left(N_1 e^{-k^2 a_1^2}+N_2 e^{-k^2 a_2^2}\right)\cos(k x_0)\nonumber\\ &+&\frac{  g_{12} N_1 N_2}{\sqrt{\pi}w} e^{-x_0^2/w^2}-\langle {\cal L}_z\rangle.
\end{eqnarray}}
using $\frac{N}{4}\frac{d^2 x_0}{dt^2}=-\frac{\partial V_{\rm eff}}{\partial x_0}$. In Fig. 1, we display variation of effective potential $V_{\rm eff}$  with the separation between the solitons keeping the velocity ($\dot{x}_0$) as constant for different values of Rabi-coupled lattice strengths. One sees that the effective potential contains of central minimum. This is basically arises due to inter-soliton interaction.  On either sides of the central minimum, we see two minima due to optical lattice.  Interestingly,  the effective potential in presence of Rabi-coupling lattice (RL) changes near the central minimum. Particularly, due to interplay between the  non-linear inter-component interaction and Rabi-coupling lattice,  shape of the potential gets modified. The height of the well decreases and local minima arises on either sides of the central minimum. The depth of local mimnimum increases  with the increase of the RL strength.  This clearly indicates that  the dynamics of soliton changes significantly with the increase of Rabi-coupling lattice strength.
\begin{figure}[h!]
\includegraphics[scale=0.40]{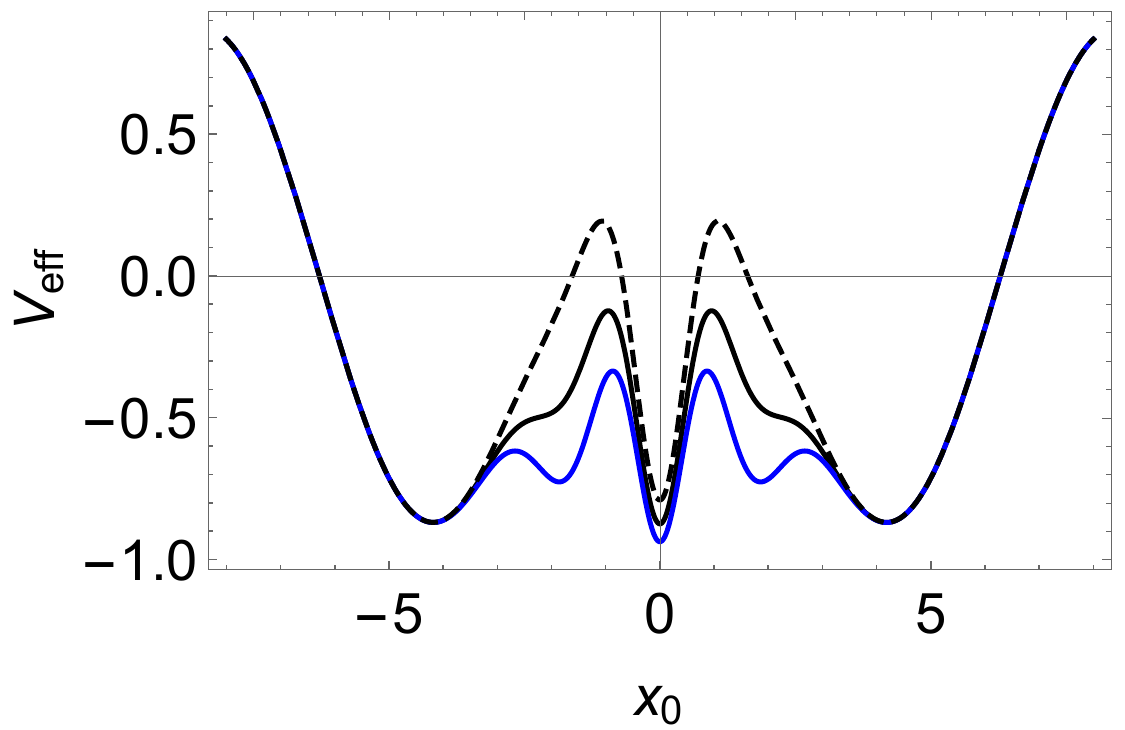}
\caption{Effective potential for different values of $\Omega$, namely, $\Omega_0=0.15$ (black dashed), $0.35$(black solid) and $0.50$(blue). Other parameters are taken as: $a_1=a_2=0.5$, $N_1=N_2=1$, $k_m=1.5$, $k=0.75$, $V_0=0.5$, $g_{12}=2$ and $\phi=\pi$. }
\label{fig1}
\end{figure} 
%%%%%%%%%%%%%%%%%%%%%%%%
%%%%%%%%%%%%%%%%%%%%%
\begin{figure}[h!]
\includegraphics[scale=0.18]{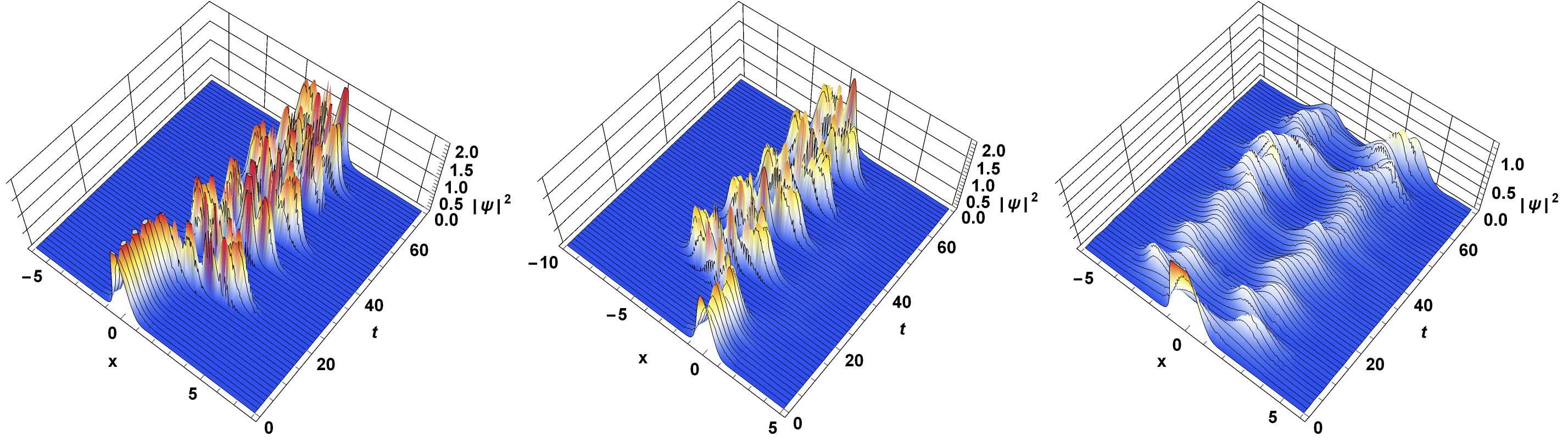}
\caption{Evolution of density profile ($|\psi|^2$) with $\Omega_0=0.15$ for $x_0=0.1$ (left), $x_0=0.4$(middle) and $x_0=0.75$(right panel). In all the panels we take $V_0=0.5$, $\Omega_0=0.35$, $g=2.51$, $g_{12}=2.0$, $N_1=N_2=1$, $a_1=a_2=0.5$,$k=0.75$, $k_m=1.5$ and $k_s=0.1$. }
\label{fig2}
\end{figure} 
%%%%%%%%%%%%%%%%%%%%%%%%%%%%%%%%
%%%%%%%%%%%%%%%%%%%%%
\begin{figure}[h!]
\includegraphics[scale=0.21]{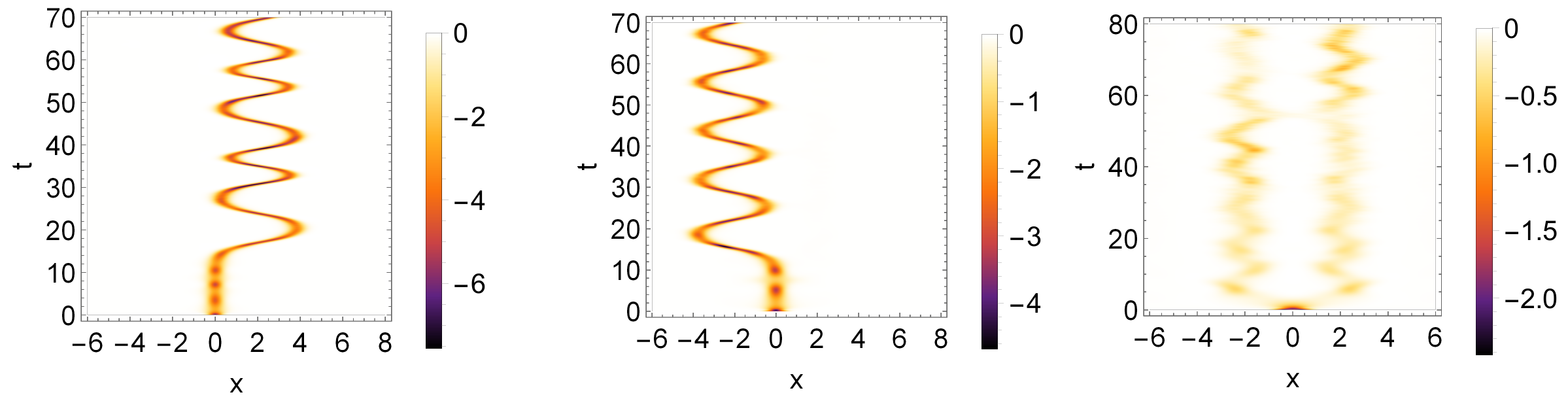}
\caption{Time evolution of  effective energy $V_d(x,t)$  for the values of parameters given  in Fig.2. Here the trajectory gives minimum value of $V_d$.}
\label{fig3}
\end{figure} 
%%%%%%%%%%%%%%%%%%%%%%
%%%%%%%%%%%%%%%%%%%%%%%%%%%%%5
\section{Numerical  simulation of soliton-components' dynamics}
We numerically solve the spin-orbit coupled Gross-Pitaevskii equations in (\ref{eq1}) and (\ref{eq2}) using the method of lines (MOL) \cite{mol1,mol2} for the initial condition given in Eq.(\ref{eq4}) by changing the separation between the components. In other words, we change the overlap between the solitons and allow them to evolve for different values of Rabi-coupled lattice (RL) strengths. Particularly, we consider the three cases corresponding to three different values of RL strengths ($\Omega_0$). For a relatively weaker value of RL ($\Omega_0=0.15$),  dynamics of  {total density profile of coupled solitons, $|\psi|^2=|\psi_1|^2+|\psi_2|^2$},  for different initial overlap  is shown in Fig. 2. For small initial separation, the binary matter-wave solitons merge and the total density profile initially tends to localize in the principal  minimum of  the effective potential.  However, the system does not exhibit oscillation about this minimum (zero separation) for a longer time.  The coupled soliton moves to another spatial location without changing their separation after a while. We note that this motion is regulated by the interplay between  inter-component interaction and Rabi-coupling lattice. In order to understand the dynamical behaviour of density profile, we calculate time variation of the energy $V_d(x,t)$,
\begin{eqnarray}
V_d(x,t)=\sum\left[\Omega(x) \psi_{3-j} \psi^*_j-g_{12} |\psi_j|^2|\psi_{3-j}|^2\right]
\end{eqnarray}
arising from inter-component interaction and Rabi-coupling lattice potential. The result is displayed in Fig. 3. We see that the location of minimum value of potential $V_d$ changes with time  and the total density profile follow the same path. This is quite similar to the dynamical self-trapping  of solitons  in the minimum of nonlinear optical lattice potential\cite{go3}. We examine that the dynamics is controlled initially by inter-component interaction since the $V_d$ has no minimum due to Rabi-coupling lattice only(left panel). As they move slightly in space due to optical lattice, the dynamics is determined by the Rabi-coupling lattice. We note that one can regulate the spatial location of oscillation by taking an appropriate value of initial overlap (middle panel).

The density profile  splits into two fragments if the initial overlap between the condensates is  small since the effects of inter-component interaction become weak (right panel). The splited parts of the density exhibit motion about the secondary  minimum of  $V_{\rm {eff}}$ at $x_0 \approx 4$. In this case, the dynamical potential and the potential of separation both approximately describe the same dynamics. Therefore, we see that  localization  dynamics changes with the increase of initial overlap.

\begin{figure}[h!]
\includegraphics[scale=0.18]{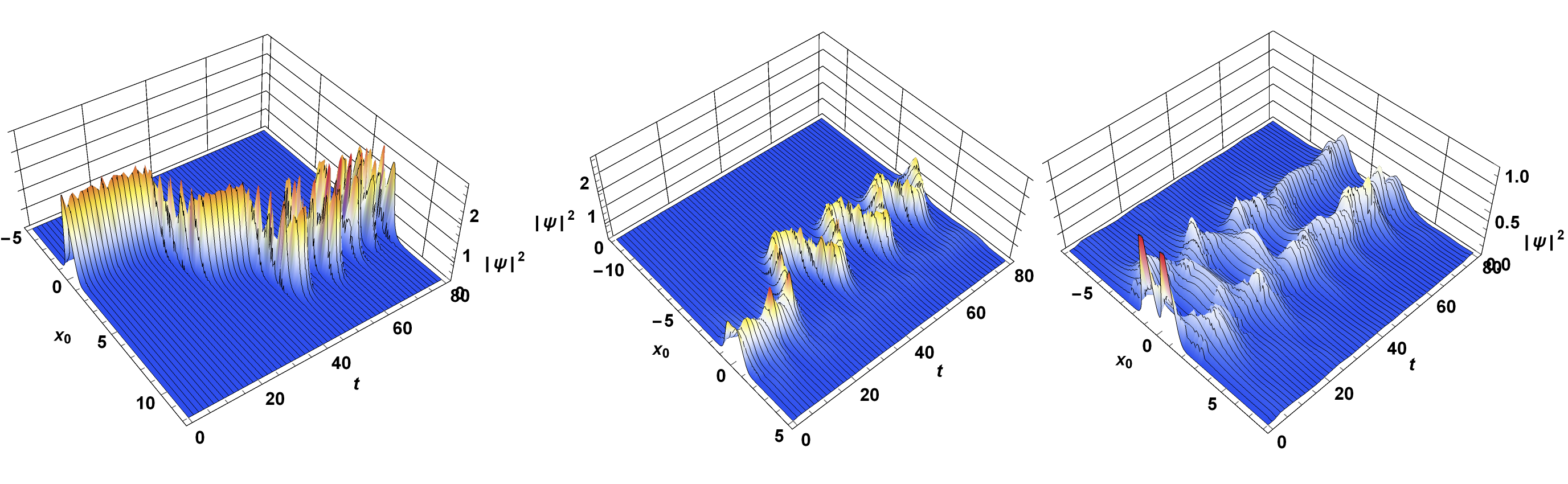}
\caption{Evolution of total density profile ($|\psi|^2$) with $\Omega_0=0.35$ for $x_0=0.2$ (left), $x_0=0.55$(middle) and $x_0=1.5$(right panel). Other parameters are kept same as those used in Fig. 2. }
\label{fig4}
\end{figure}  
%%%%%%%%%%%%%%%%%%%%%%%%%
%%%%%%%%%%%%%%%%%%%%%
\begin{figure}[h!]
\includegraphics[scale=0.2]{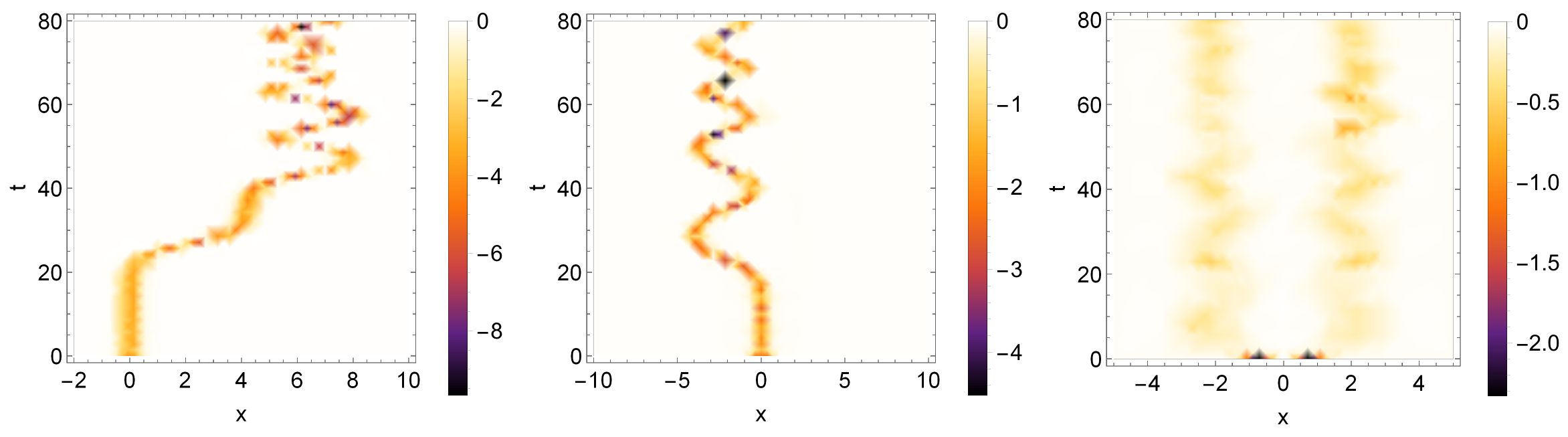}
\caption{Time evolution of effective energy $V_d(x,t)$  for the values of parameters given in Fig.4. Here the trajectory gives minimum value of $V_d$.}
\label{fig5}
\end{figure} 
%%%%%%%%%%%%%%%%%%%%%%%%%%%

If the Rabi-coupling lattice (RL) is relatively stronger than that used in Fig.3, two local minima on either side of the central minimum tend to appear in the effective potential for the separation. Thus, the dynamical behavior of  the density profile is  significantly modified from that observed for weak RL strengths. Particularly, the density profiles merge and get localized for a short time in the minimum of the separation potential (Fig. 4). After that the profile moves to another position and executes oscillation about the minimum of $V_d$ without spliting. We notice that the change of spatial position of the total density profile occurs due to change of RL strength (Fig. 5).

%%%%%%%%%%%%%%%%%%%%%%
\begin{figure}[h!]
\includegraphics[scale=0.2]{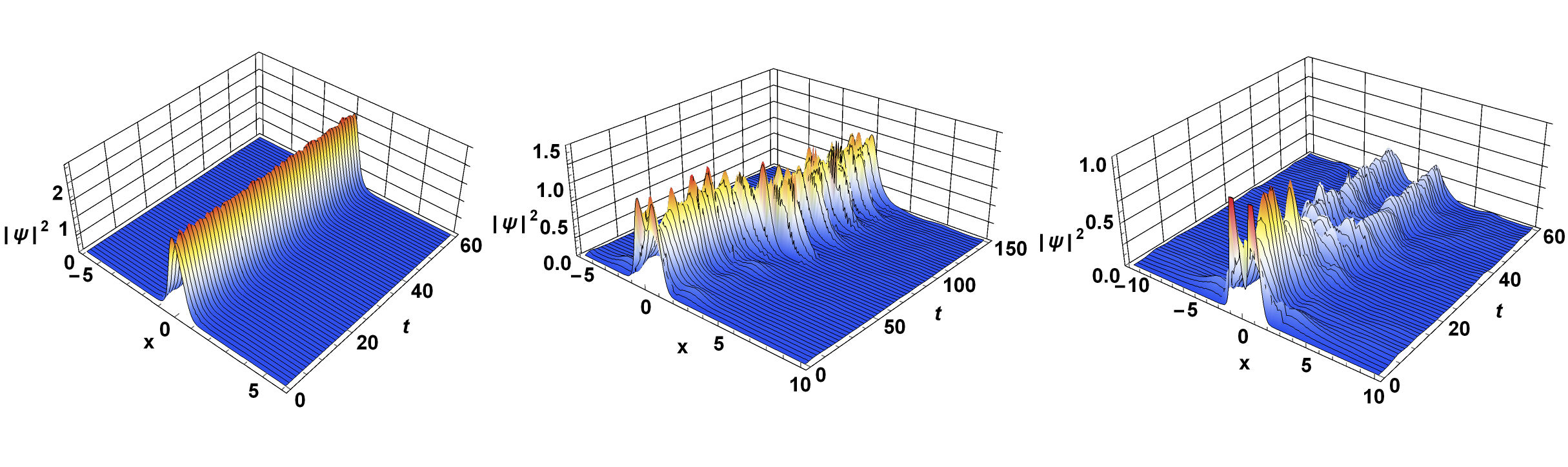}
\caption{Evolution of density profile ($|\psi|^2$) in presence of strong Rabi lattice with $\Omega_0=0.50$ for $x_0=0.2$ (left panel), $1.02$(middle) and $1.25$(right panel). In all the panels we take $V_0=0.5$, $\Omega_0=0.35$, $g=2.51$, $g_{12}=2.0$, $N_1=N_2=1$, $a_1=a_2=0.5$, $k=0.75$, $k_m=1.5$ and $k_s=0.1$.}
\label{fig6}
\end{figure} 
%%%%%%%%%%%%%%%%%%%%%%%%%%%%%
%%%%%%%%%%%%%%%%%%%%%
\begin{figure}[h!]
\includegraphics[scale=0.23]{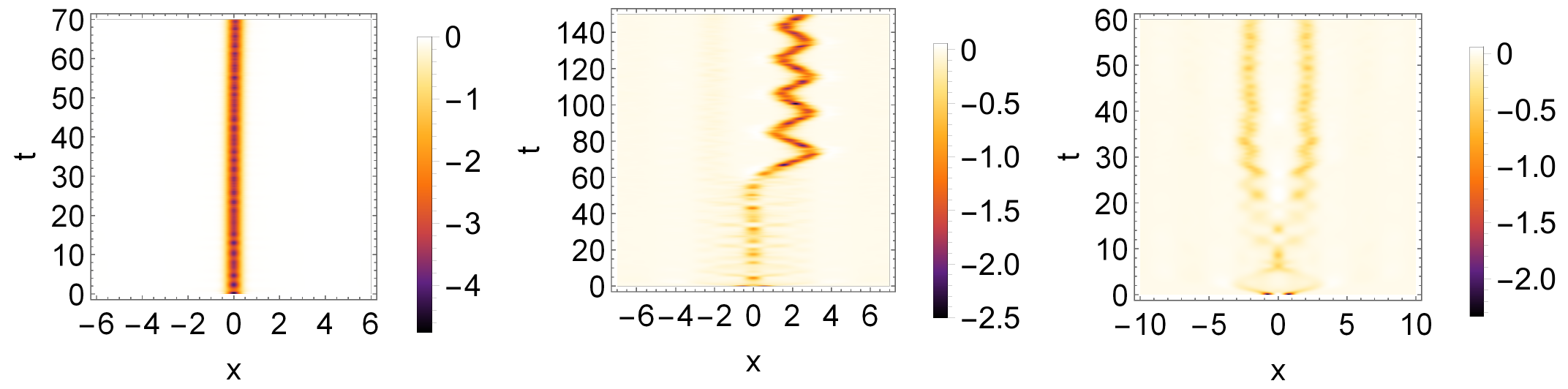}
\caption{Time evolution of  effective  energy $V_d(x,t)$  for the values of parameters given in Fig.6. Here the trajectory gives minimum value of $V_d$. }
\label{fig7}
\end{figure} 
%%%%%%%%%%%%%%%%%%%%%%%%%%%

In the case of stronger Rabi-coupling lattice,  depth of central minimum of the separation potential increases. Thus, for a larger initial overlap between the components, the total density profile  of the binary solitons is strongly localized there and do not move with time (Fig.4). With the reduction of initial overlap, the inter-component interaction reduces and the RL tends to dominates. This results movement of the profile along  the path of minimum of $V_d$ (Fig. 7). Finally, the motion is settled down at the place where RL is dominating. For very small initial spatial overlap, the inter-component interaction can not hold them together and density profile gets splitted. In this case, the density of condensate tends to localize in the local minima of the  the effective potential.

%%%%%%%%%%%%%%%%%%%%%%
\begin{figure}[h!]
\includegraphics[scale=0.28]{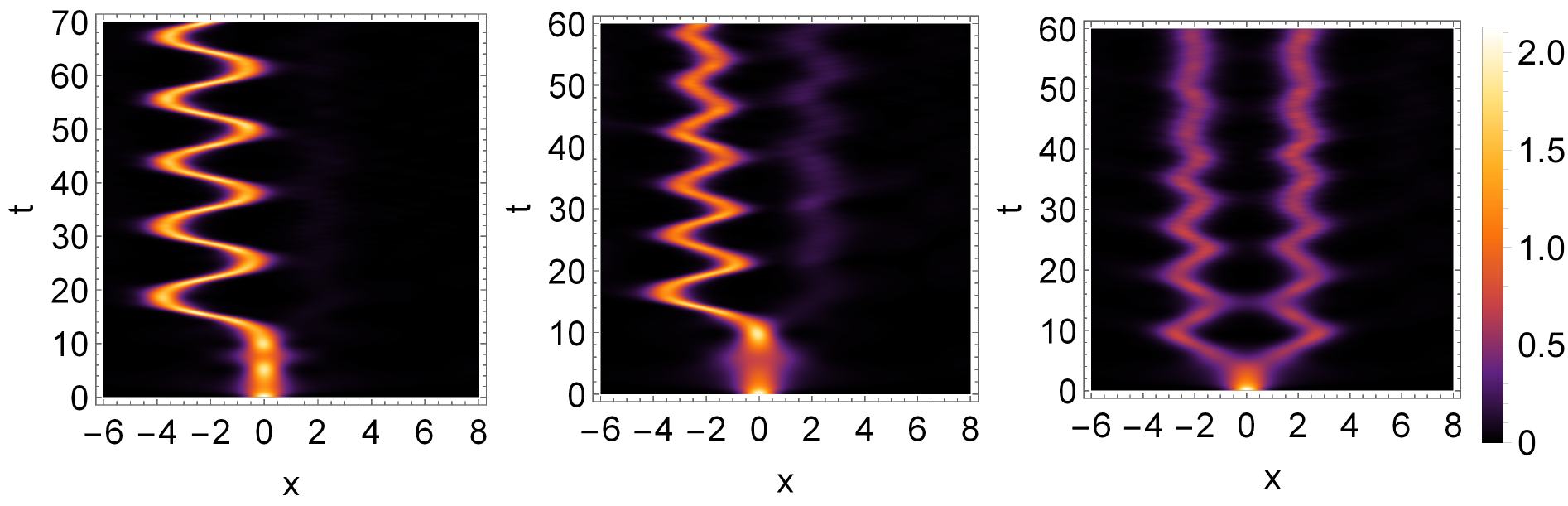}
\caption{Evolution of total density profile ($|\psi|^2$) in presence of Rabi-coupling lattice with  $x_0=0.4$ for $\Omega_0=0.1$(left panel), $0.065$(middle) and $0.05$(right panel). In all the panels we take $V_0=0.5$, $g=2.51$, $g_{12}=2.0$, $N_1=N_2=1$, $a_1=a_2=0.5$, $k=0.75$, $k_m=1.5$ and $k_s=0.1$.}
\label{fig8}
\end{figure} 
%%%%%%%%%%%%%%%%%%%%%%%%%%%%%

{We see that the spin-components  splits with change of their initial environment ($x_0$) in the effective potential for fixed value Rabi-coupling lattice (RL) strength. It is quite interesting to check if there exist any threshold value of RL strength for a fixed initial environment in the effective potential $V_{\rm eff}(x_0)$. In Fig. 8, we show the evolution of total density profile $|\psi|^2=|\psi_1|^2+|\psi_2|^2$ for different $\Omega_0$ values. The total density remains unsplitted if $\Omega_0 > 0.065$ (left panel). It starts to split when $\Omega_0 \approx 0.065$ (middle panel). The density is completely splitted for $\Omega_0=0.05$. The threshold value for splitting, however, changes with change of RL wave number. Particularly, the threshold value  of $\Omega_0$ for splitting increases with the decrease of RL wave number and vice-versa.  }

\section{conclusions}
Spin-independent optical lattice serves as an artificial periodic potential and it is  found useful to simulate different phenomenon of condensed matter physics. It is found also useful to control dynamics of matter-waves solitons in Bose-Einstein condensates. Bright solitons in spin-orbit coupled Bose-Einstein condensates (SOC-BECs) generally  suffers from the loss Gallelian  invariance which results shape change as the soliton propagates. In this work, we investigate the role of Rabi-coupling lattice (RL) on the dynamical behaviour of the matter-wave bright solitons.

We have constructed an effective potential for the separation of center of mass of two solitons in spin-orbit coupled Bose-Einstein condensates. The effective potential is very sensitive with the change of Rabi-coupling lattice (RL) strength. Particularly, depth of the central minimum increases with the increase of  RL strength and also additional local minima arise near the central minimum. Thus RL  can generate additional local trapping centers.

We have studied numerically the dynamical behavior of the density profiles for different RL strength by changing the initial overlap of soliton components. We have demonstrated that the dynamics of the density profile is controlled by the interplay between initial overlap and Rabi-coupling  lattice.  For smaller values of RL strength, density profiles of the component solitons merge due  to stronger effective inter-component interaction and the total density profile tries to be localized for a shorter time. Due to the acceleration caused by the lattice the profile gets shifted to another place where it executes oscillation about minimum resulting from Rabi-coupling lattice. If the overlap between the condensates is small then inter-component interaction can not hold them and the total density splitted  and then fragments execute oscillatory motion about the local minima.  For a stronger value of RL, the total density profile highly localized if the overlap between the condensates is large. However, it is splitted  and the fragments localized in the local minima of the effective potential if the spatial overlap is small. {We have examined that there exists a threshold value RL strength for the occurrence of splitting of spin-components and this threshold depends on the RL wave number.}  Thus, we conclude by noting that the RL can be properly tuned to realized, merging, splitting and localizing dynamics of soliton pair in  spatially coherent spin-orbit coupled Bose-Einstein condensates. 

%%%%%%%%%%%%%%%%%%%%%%%%%%%%%
\section*{Acknowledgements}
S. Sultana would like to thank  `West Bengal Higher Education Department' for providing Swami Vivekananda Merit Cum Means Scholarship with F. No. WBP201637916084.
%%%%%%%%%%%%%%%%%%%%%%%% 

\end{document}